\newcommand{\be}{\begin{eqnarray}}
\newcommand{\ee}{\end{eqnarray}}

\newcommand{\p}[1]{(\ref{#1})}

\renewcommand\={\ =\ }

\documentclass[12pt]{article}

\usepackage{amsmath,amssymb}
\usepackage{graphicx}
\usepackage{cite}
\usepackage{bm}
\begin{document}

\renewcommand{\thefootnote}{\fnsymbol{footnote}}

\vskip 15mm

\begin{center}

{\Large Cryptoreality of nonanticommutative Hamiltonians}

\vskip 4ex

E.A. \textsc{Ivanov}$\,^{1}$,
A.V. \textsc{Smilga}\,$^{2}$,

\vskip 3ex

$^{1}\,$\textit{ Bogoliubov  Laboratory of Theoretical Physics, JINR, 141980 Dubna, Moscow Region,
Russia}
\\
\texttt{eivanov@theor.jinr.ru},\\[3ex]
$^{2}\,$\textit{SUBATECH, Universit\'e de
Nantes,  4 rue Alfred Kastler, BP 20722, Nantes  44307, France
\footnote{On leave of absence from ITEP, Moscow, Russia.}}
\\
\texttt{smilga@subatech.in2p3.fr}
\end{center}

\vskip 5ex

\begin{abstract}
\noindent We note that, though nonanticommutative (NAC) deformations of Minkowski supersymmetric
theories do not respect the  reality condition and seem to lead to non-Hermitian Hamiltonians $H$, the
latter belong to the class of ``cryptoreal'' Hamiltonians considered recently by Bender and
collaborators. They can be made manifestly Hermitian via the similarity transformation
$H \to e^R H e^{-R}$ with a properly chosen $R$. The deformed model enjoys {\it the same} supersymmetry
algebra as the undeformed one, though being realized differently on the involved canonical variables.
Besides quantum-mechanical models, we treat, along similar lines, some NAC deformed field models in
$4D$ Minkowski space.
\end{abstract}

\renewcommand{\thefootnote}{\arabic{footnote}}
\setcounter{footnote}0
\setcounter{page}{1}

\section{Introduction}
Supersymmetric models with nonanticommutative (NAC) deformations \cite{Seiberg} have
recently attracted  a considerable
interest. The main idea is that the odd superspace coordinates $\theta^\alpha$ and
$\bar \theta^{\dot \alpha}$
are not treated as strictly anticommuting anymore, but involve  non-vanishing
anticommutators \cite{Nonc} \footnote{In other words, the original Grassmann algebra of the odd coordinates
is deformed into a {\it Clifford} algebra.}.
In  original Seiberg's paper and in many subsequent works (see e.g. \cite{Euclid,Ito} and
references therein), the deformation
is performed in Euclidean
rather than Minkowski space-time. The reason is that in Minkowski space it seems  impossible
to preserve
both supersymmetry and reality of the action after deformation, still retaining simple properties of
the corresponding $\star$-product (e.g., associativity and nilpotency) \cite{Mink}.
As discussed in \cite{Seiberg}, Euclidean
NAC theories are of interest in stringy perspectives \footnote{The stringy origin
of such deformations \cite{StrOr}
was actually the main motivation of their consideration in \cite{Seiberg} (see also \cite{N2,Ito}).}.
An interesting question
is whether NAC theories are meaningful by themselves, leaving aside the issue
of their relationships
with string theory.
In other words --- whether it is possible
to consistently define them in Minkowski
space, introduce a Hamiltonian with real spectrum and find
a unitary evolution operator.

We argue that the answer to this question is positive. Our consideration is mostly based on the analysis of
an interesting 1--dimensional NAC model constructed in a recent paper of Aldrovandi
and Schaposnik \cite{shap}.
In that work, NAC deformations of the conventional Witten's supersymmetric quantum mechanics (SQM)
model \cite{Witten} were studied in the chiral basis. In this case,
the deformation operator commutes with the supercharge $Q$, but does not commute with $\bar Q$. However,
Aldrovandi and Schaposnik noticed the presence of the second supercharge $\bar {\cal Q}$ that commutes with
the Hamiltonian. On the other hand, $Q$ and    $\bar {\cal Q}$ seem not to be Hermitian conjugate to each other
and the deformed Hamiltonian also seemingly lacks the Hermitian property.

Our key observation is that, in spite of having a complex appearance, this Hamiltonian is actually
Hermitian in disguise. One can call it ``crypto-Hermitian'' (or ``cryptoreal''). It belongs to the
class of Hamiltonians studied recently by Bender and collaborators \cite{Bender}. The simplest example is
 \be
\label{ix3}
H \ =\ \frac {p^2 + x^2}2 + igx^3\,.
 \ee
In spite of the manifestly complex potential, it is possible to endow the Hamiltonian (\ref{ix3})
with a properly defined Hilbert space such that the spectrum of $H$ is real. The clearest way to see
this is to observe
the existence of the operator $R$ such that the conjugated Hamiltonian
  \be
\label{conjug}
  \tilde H \ = \ e^{R} H e^{-R}
  \ee
 is manifestly
self-adjoint \cite{turok}. The explicit form of $R$ for the Hamiltonian (\ref{ix3}) is
\footnote{Actually, what is written here is the Weyl symbol of the operator $R$. The expression
for a contribution
to the quantum operator corresponding to a monomial $\sim p^n x^n$ in its Weyl symbol is a properly
symmetrized structure, $px \to (1/2)(\hat p x + x \hat p)$, $x^2 p
\to (1/3)( x^2 \hat p + \hat p x^2 + x\hat p x)\,$, etc.}
 \be
\label{Rix3}
  R = g\left( \frac 23 p^3 + x^2 p \right) - g^3 \left( \frac {64}{15} p^5 + \frac {20}3 p^3 x^2 +
4px^4 - 6p \right) + O(g^5)\ .
 \ee
The rotated Hamiltonian is
 \be
 \label{hix3}
\tilde H \ =\ \frac{p^2 + x^2}2 + g^2 \left( 3p^2x^2 + \frac {3x^4}2 - \frac 12 \right) + O(g^4)\ .
 \ee
The (real) spectrum of $\tilde H$ (and $H$) can be found to any order in g in the perturbation theory,
and also non-perturbatively.

 We will see that in  the case of the Aldrovandi-Schaposnik Hamiltonian, there also exists
the operator $R$  making the Hamiltonian Hermitian. The rotated supercharges $e^R Qe^{-R}$ and
$ e^R \bar {\cal Q} e^{-R}$ are Hermitian-conjugated.

We start in Section 2 by constructing the operator $R$ for certain non-supersymmetric
Hamiltonians. In particular, we discuss holomorphic deformations
(adding to the Hamiltonian a holomorphic function of a complex dynamic variable). In Section 3, we present
the Aldrovandi-Schaposnik model, find  the corresponding operator $R$, as well as the rotated Hamiltonian and supercharges.
Also we briefly consider a NAC deformation of the SQM model with two sorts of chiral supermultiplets.
In Section 4, we discuss possible generalizations to field theory.

\section{Cryptoreality: some comments}
 \begin{itemize}

\item
First, about the term ``cryptoreality''. In the original papers \cite{Bender}, the Hermiticity of the
Hamiltonian (\ref{ix3}) and its relatives was deduced
from a certain special symmetry of this Hamiltonian, the ${\cal PT}$-symmetry. Indeed, the Hamiltonian
(\ref{ix3}) is invariant
with respect to the combination of the parity transformation (which changes the sign of $x$)
and the time reversal
transformation (which changes $i$ to $-i$). The ${\cal PT}$-symmetry of the Hamiltonian might be
a sufficient condition for the
existence of the operator $R$ such that the conjugated Hamiltonian (\ref{conjug}) is manifestly Hermitian,
but, as we will see later, it is not a necessary condition. In Ref.\cite{turok},
the term ``pseudo-Hermiticity''
was used. To our mind, however, there is nothing ``pseudo'' about it, the Hamiltonian (\ref{ix3}) is simply
Hermitian (in the properly defined Hilbert space), but its Hermiticity is  hidden, not immediately
obvious. That is why the term ``crypto-Hermiticity'' (or ``cryptoreality'') seems to us somewhat more
appropriate.

\item
The conjugation (\ref{conjug}) acts upon all operators including the operators $p,x$. The Weyl symbols of the
transformed operators $p', x'$ are
  \be
 \label{p,x}
p' &=& p + 2igxp + g^2(2p^3 - px^2) + \ldots \nonumber \\
x' &=& x -ig(x^2 + 2p^2) - g^2 (x^3-2xp^2) + \ldots\,.
  \ee
One can actually obtain the expression (\ref{hix3}) for the Weyl symbol of the rotated Hamiltonian by simply
expressing $H$ in terms of $p', \, x'$.
The commutator $[p,x]$ is not changed after conjugation, that means that the {\it Moyal bracket}
$\{p', x'\}_{M.B.}$ is equal to one. The Moyal bracket is defined as \cite{Moyal}
 \be
\label{Moyal}
\{A,B\}_{M.B.} \ =\ 2 \left. \sin \left[ \frac 12 \left( \frac {\partial^2}{\partial p \partial X} -
 \frac {\partial^2}{\partial P \partial x} \right) \right] A(p,x) B(P,X) \right|_{p=P, x=X}
  \ee
The expansion starts with the   Poisson bracket, but, generically, there are also higher terms.
 In particular, $\{p', x'\}_{M.B.}$ differs from
$\{p', x'\}_{P.B.}$ by the terms of order $\sim g^4$ and higher. But that means that (\ref{p,x}) {\it is}
not a canonical transformation. And this means that the {\it classical} dynamics of $H(p,x)$ and $H(p',x')$
are different. The quantum dynamics of the original and conjugated Hamiltonians is, however,
the same.

\item
One can rotate away not only imaginary pieces in the potential, but also other unfriendly looking terms in the
Hamiltonian. For example, one can consider the Hamiltonian
 \be
\label{Hx3}
H \ =\ \frac {p^2 + x^2}2 + gx^3
 \ee
and conjugate it with the operator $R$ coinciding with the expression in
 Eq.(\ref{Rix3}) multiplied by the factor $-i$. The conjugated Hamiltonian
coincides with (\ref{hix3}), with the sign of $g^2$ being reversed. The spectrum of the Hamiltonian (\ref{Hx3})
can be found by the same
token as for the Hamiltonian (\ref{ix3}). Actually, an exact mapping relating the system (\ref{Hx3}) to
the system (\ref{ix3}) exists. Indeed, for any eigenfunction $\Psi_n(x)$ of the Hamiltonian (\ref{ix3})
with eigenvalue $E_n$,
the function $\Psi_n(-ix)$ is an eigenfunction of the Hamiltonian (\ref{Hx3}) with the eigenvalue $-E_n$.

The appearance of complex values of $x$ may be somewhat unusual, but it is actually an inherent
feature of the crypto-Hermitian systems.
The eigenfunctions of the Hamiltonian are required to behave well (be not  singular and  die  out for
large absolute values of the argument)
in a certain domain in the complex $x$-plane that might or might not include the real axis \cite{Bender}.
The relevant domains for the Hamiltonians
(\ref{ix3}) and (\ref{Hx3}) are shown in Fig. \ref{oblastx3}. One is rotated with respect to the other by
the angle $\pi/2$.

 \begin{figure}[h]
   \begin{center}
 \includegraphics[width=4.0in]{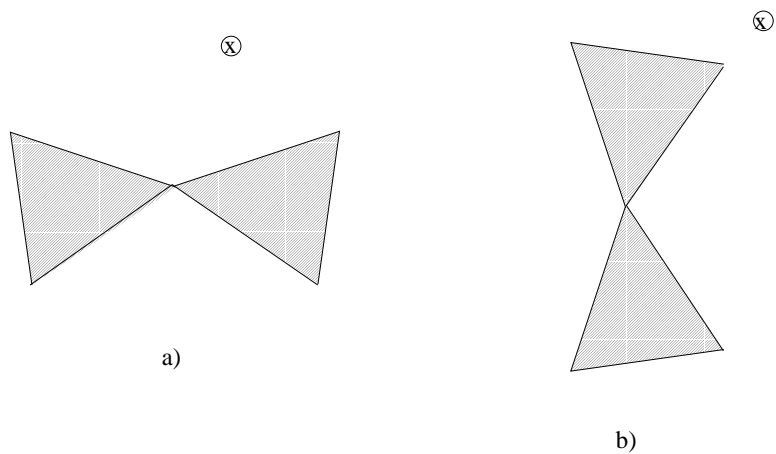}
    \end{center}
\caption{Wave functions for {\it a)} the Hamiltonian (\ref{ix3}) and {\it b)}
  Hamiltonian (\ref{Hx3}) asymptotically die out in the dashed sectors.}
\label{oblastx3}
\end{figure}

Another unusual feature of the Hamiltonian (\ref{Hx3}) is the absence of the ground state -
the state
with the lowest energy.
In this case, the spectrum has an upper rather than lower bound. But the overall sign of energy
is in fact a matter of book-keeping.
For all physical purposes,
the dynamics of the Hamiltonian (\ref{ix3}) in the region in Fig.\ref{oblastx3}a and
the dynamics of the Hamiltonian
(\ref{Hx3}) in the region in Fig.\ref{oblastx3}b are equivalent.

Consider now the Hamiltonian
 \be
\label{Hz}
H \=\ \bar \pi \pi + \bar z z + gz^3 \, .
 \ee
 Remarkably, by conjugating it with the operator
 \be
\label{Rz}
R = -ig\left( \bar \pi z^2 + \frac 23 \bar \pi^3 \right),
 \ee
one can rotate away the cubic term in the potential {\it without trace} such that the conjugated Hamiltonian
$H' = \bar \pi' \pi' + \bar z' z' + g z'{}^3$ is simply $\bar \pi \pi + \bar z z$. Hence the spectrum of
the Hamiltonian (\ref{Hz}) coincides with the
spectrum of a 2-dimensional oscillator, $E_{n,m} = 1 + n + m\,$.  The wave functions of the original
Hamiltonian (\ref{Hz}) are obtained from the
oscillator wave functions
by conjugation $\Psi = e^{-R} \tilde \Psi$. For example, the ground state wave function is
 \be
\label{ground}
\Psi_0 \ \sim \ \exp\left\{ - \frac {gz^3}3 - \bar z z \right \}.
 \ee
It decays exponentially in the three sectors in the complex plane of $z$ shown in Fig. \ref{oblastz3}, and
the Hilbert space where the crypto-Hermitian
Hamiltonian (\ref{Hz}) is well defined is  formed by the functions sharing this property.

 \begin{figure}[h]
   \begin{center}
 \includegraphics[width=2.0in]{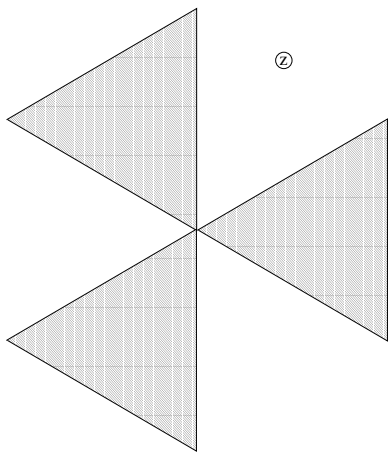}
    \end{center}
\caption{The same for the Hamiltonian (\ref{Hz}).}
\label{oblastz3}
\end{figure}

 By the same token, one can rotate away without trace {\it any} holomorphic term in the potential.
 For example,
for the Hamiltonian
$  \bar \pi \pi + \bar z z + gz^5 $, this is done with the operator
 $$
 R = -ig\left( \bar \pi z^4 + \frac 43 \bar \pi^3 z^2 + \frac 8{15} \bar \pi^5 \right).
 $$
Generally, the operator rotating away the term $gz^N$ in the potential has the form
  $$
 R_N = -ig \bar \pi z^{N-1} f_N\left( \frac {\bar \pi}{z} \right),
 $$
with $f_N(r)$ satisfying the equation
 \be
\label{eqf}
[1 - r^2(N-1)] f_N + r (1+r^2) f'_N = 1\ .
 \ee
When  $N$ is odd, the solution represents a polynomial. For even $N$, it is more complicated. For example,
 \be
\label{f2}
f_2(r) = \frac 12 \left[\frac {1+r^2}r \arctan r + 1 \right].
 \ee

\item
 Cryptoreal Hamiltonians for the systems with continuum number of degrees of freedom also exist.
 Bender, Brody, and Jones
found the proper conjugation operator for the system described by the Lagrangian \cite{Bender}
 \be
 \label{Lfi3}
 {\cal L} \ =\  \frac 12 (\partial_\mu \phi )^2 - \frac {\mu^2 \phi^2}  2 - ig \phi^3\ ,
\ee
$\phi$ is a real scalar field.
In the lowest order in $g$, it is given by a nonlocal expression
 \be
\label{Rphi3}
 R \ =\ \int \!\!\int\!\!\int d{\bf x}  d{\bf y}  d{\bf z}
 \left[ M_{ {\bf x}  {\bf y}  {\bf z} } p_{\bf x}  p_{\bf y}  p_{\bf y}
+ N_{ {\bf x}  {\bf y}  {\bf z} } \phi_{\bf x} \phi_{\bf y} p_{\bf z} \right],
  \ee
where $p_{\bf x}$ are canonical momenta, $p_{\bf x} = -i\partial /\partial \phi_{\bf x}\,$,
 and the kernels $ M_{ {\bf x}  {\bf y}  {\bf z} },  N_{ {\bf x}  {\bf y}  {\bf z} }$
 have a complicated, but explicit form.

We want to notice that the system of the complex scalar field $\varphi$ with the interaction Hamiltonian
$\sim \varphi^3$ is also cryptoreal,
and the corresponding conjugation operator is given, again, by the expression (\ref{Rphi3}) with
$\bar \pi_{\bf x} = -i \partial /\partial \bar
\varphi_{\bf x}$ being substituted for $p_{\bf x}$. This operator rotates the interaction term away
without trace by the same token as the
operator (\ref{Rz}) rotates it away in the QM case.

Actually, the pattern is quite general. Any holomorphic interaction term can be entirely rotated away
simply because the proper conjugation operator
$R$ involves in this case only the momentum operators $\bar \pi_{\bf x}$ rather than  $\pi_{\bf x}\,$,
and $\bar \partial f = 0$ for holomorphic functions.

\item
Finally, let us reproduce here the arguments of \cite{Bender} displaying the reality of the spectrum
of a ${\cal PT}$-symmetric  Hamiltonian.
The operator ${\cal PT}$ commutes with the Hamiltonian, and it is reasonable
to assume that a basis of the states representing the eigenstates
of both ${\cal PT}$ and $H$ can be chosen \footnote{Were ${\cal PT}$ a linear operator, it would be
trivial, but ${\cal PT}$ involves complex conjugation and is not linear. Hence, the existence of such
basis is, indeed,
an {\it assumption} and  the reasoning given here cannot be regarded as a formal proof.}.
  Let $\Psi$ be an
eigenstate of both ${\cal PT}$ and $H$,
 \be
{\cal PT} \, \Psi \ =\ \lambda \Psi,\ \ \ \ \ \ \ \ \ H\, \Psi \ =\ E\Psi\ .
 \ee
Applying the operator ${\cal PT}$ to the second equality and using $[{\cal PT},H] = 0 $ and
$  {\cal PT} (E\Psi) = E^* {\cal PT}(\Psi)\,$, we conclude that
$E = E^*$ {\it Q.E.D.} Note also that applying ${\cal PT}$ to the first equality and
using $({\cal PT})^2 = 1$, one can show that
$\lambda \lambda^* = 1$ and hence $\lambda = e^{i\alpha}$. By going from $\Psi$
to  $\Psi e^{-i\alpha/2}$, one can set $\lambda = 1\,$.

The norm of some eigenstates may happen to be negative.  However, this can be mended \cite{Bender}
if redefining  inner product by including
in its definition the action of the  ``charge conjugation'' operator ${\cal C}$  that commutes
with both $H$ and ${\cal PT}$ and is defined as
 \be
 \label{C}
{\cal C}(x,y) \ =\ \sum_n \Psi_n(x) \Psi^*_n(y)\ .
 \ee
The operator ${\cal C}$ is in fact closely related to the operator $R$ rotating the Hamiltonian
to the manifestly Hermitian form,
as discussed above,
 \be
{\cal C}\ =\ e^{-2R}{\cal P}\, .
 \ee

\end{itemize}

\section{Aldrovandi-Schaposnik model}

The simplest SQM model \cite{Witten} involves a real supervariable
\be
\label{X}
X(\theta, \bar \theta, t) \  = \ x(t) + \theta \psi(t)
+ \bar \psi(t) \bar \theta  + \theta \bar \theta F(t)\ .
 \ee
The action is
 \be
\label{LSQM}
 S \ =\ -\int dt \, d^2\theta \left[ \frac 12 (DX) (\bar D X)  + V(X) \right],
 \ee
with the convention $\int d^2\theta \, \theta \bar\theta = 1\,$. Here $V(X)$ is the superpotential
and $D, \bar D$ are covariant derivatives. Bearing in mind the
deformation coming soon, we will choose their left chiral basis representation
 \be
\label{DDbar}
 D \ =\ \frac \partial{\partial \theta}  - 2i  \bar\theta \frac \partial {\partial t} \, ,\ \ \ \
 \ \ \bar D \ =\
- \frac \partial{\partial \bar \theta} \ .
 \ee
Here $t = \tau -i\theta\bar\theta\,$ and $\tau$ is the real time coordinate of the central basis.
Asymmetry between $D$ and $\bar D$ makes the Lagrangian following from (\ref{LSQM}) complex,
 \be
\label{Lnedef}
 L \ =\ -i\dot x F - \frac{\partial V(x)} {\partial x} F + \frac 12 F^2 + i \bar \psi \dot \psi +
\frac {\partial^2 V(x)}{\partial x^2}
\bar \psi \psi \, ,
 \ee
but one can easily make it real, rewriting it in terms  of $\tilde F = F - i\dot x$
and subtracting a total derivative.
This corresponds to going
over to the central basis from the chiral one.

The deformation is introduced by postulating non-vanishing anticommutators
 \be
\label{deform}
\{\theta, \theta\} = C,\ \ \ \ \{\bar\theta, \bar\theta\} = \bar C, \ \ \ \ \ \{\theta, \bar \theta\}
= \tilde C \, .
  \ee
 The deformed action  involves star products,
   \be
\label{Ldeform}
 S \ =\ -\int dt \, d^2\theta \left[ \frac 12 (D \star X)\star (\bar D \star X)
 +  V_{\star}(X) \right],
 \ee
where
 \be
\label{star}
 X\star Y \ =\  \left. \exp \left\{ - \frac C2 \frac {\partial^2 }{\partial \theta_1 \partial \theta_2}
  - \frac {\bar C}2
\frac {\partial^2}{\partial \bar \theta_1 \partial \bar \theta_2 }
  - \frac {\tilde C}2 \left(
\frac {\partial^2}{\partial  \theta_1 \partial  \bar \theta_2 } +
\frac {\partial^2}{\partial  \bar \theta_1 \partial   \theta_2 } \right) \right\} X(1) Y(2) \right|_{1=2}
 \ee
and $V_{\star}(X)$ is obtained from $V(X) = \sum_n c_n X^n$ by substituting $X^2 \to X_\star^2 \equiv
X\star X,\ X^3 \to X_\star^3 \equiv X\star X\star X$, etc
in its Taylor expansion. The star product in \p{Ldeform} just ensures the Weyl ordering of any
product of the $\theta $ monomials such that
$$ \theta \star\theta = \frac C2,\ \ \ \bar \theta \star \bar \theta =  \frac {\bar C}2, \ \ \
\theta \star \bar\theta =
\theta \bar\theta + \frac {\tilde C}2, \ \ \ \bar \theta \star \theta =
\bar \theta \theta + \frac {\tilde C}2\ ,$$
in accordance with the basic relation \p{deform}. The star product is associative.

The component expression for the deformed Lagrangian is the same as in Eq. (\ref{Lnedef}), with $V(x)$
being substituted by \cite{Alv,shap}
 \be
\label{Vtildint}
 \tilde V(x, F) \ =\ \int_{-1/2}^{1/2} d\xi \, V(x + \xi c F)\ ,
 \ee
where
 \be
\label{c2}
c^2 =  \tilde C^2 - C\bar C
 \ee
 is the relevant deformation parameter \footnote{ The relation (\ref{Vtildint}) can be easily derived by
keeping the term $\propto \theta \bar\theta$ in the products $X_\star^n$, with using
associativity and the identity
$ (\theta \bar\theta) \star (\theta \bar\theta) = c^2/4$. Note the correct sign of $c^2$ in \p{c2} as compared
to the wrong one in the definition of $c^2$ in \cite{shap}.}.
    If $\bar C$ is conjugate to $C$ and $\tilde C$ is real,
 $c^2$ is also real.
Note, however, that one may, generally speaking, lift the condition that $\theta$ and $\bar \theta$
are conjugate to each other, in which
case $C,\bar C$ and $\tilde C$ can take arbitrary values. We still require the reality of $c^2$.
The crypto-Hermiticity of the deformed Hamiltonian discussed below is fulfilled under this condition.

 In the simplest nontrivial case,
$V(X) = \lambda X^3/3$,
 \be
\label{Vtild}
 \tilde V(x, F) \ =\ \frac {\lambda x^3}3 + \frac {\lambda c^2xF^2}{12} \ .
 \ee
The corresponding canonical Hamiltonian is
 \be
\label{Hamdef}
H \ =\ \frac {p^2}2 + i \frac {\partial \tilde V}{\partial x} p
-  \frac {\partial^2 \tilde V}{\partial x^2} \bar \psi \psi\,,
 \ee
with $p = -iF$. The deformed Lagrangian and Hamiltonian look inherently complex. Obviously, the complexities
now cannot be removed by simply
going from the chiral to the central basis.

In the chiral basis, the supercharges are represented by the following superspace differential operators,
\be
\label{QQbar}
 Q \ =\ \frac \partial{\partial \theta} \, ,\ \ \ \ \ \ \bar Q \ =\
- \frac \partial{\partial \bar \theta}  - 2i  \theta \frac \partial {\partial t}\ .
 \ee
 Note that the star product operator (\ref{star}) still commutes with $Q$ (in  other words, the Leibnitz rule
$Q\star(X\star Y) = (Q\star X)\star Y + X\star (Q\star Y)$ still holds), but not with $\bar Q$.
That means that the deformed
action (\ref{Ldeform})
is still invariant with respect to the supersymmetry transformations generated by $Q$, but not $\bar Q$.
The $Q$-invariance implies the existence of the conserved N\"other supercharge whose component phase space
expression is simply
 \be
\label{Qcomp}
 Q \ =\ \psi p\ .
 \ee
As was observed in \cite{shap}, there is another Grassmann-odd operator commuting with the Hamiltonian.
It reads
 \be
\label{Qbarcomp}
 \bar{\cal Q} \ =\ \bar \psi \left( p + 2i \frac{\partial \tilde V}{\partial x} \right).
 \ee
The standard SUSY algebra
\be
\label{algebra}
 Q^2 = \bar {\cal Q}^2 = 0,\ \ \ \ \ \ \ \{Q, \bar{\cal Q} \} = 2H
 \ee
holds, but,
naively, $\bar{\cal Q}$ is not adjoint to $Q$ and $H$ is not Hermitian.

Let us show now that the Hamiltonian (\ref{Hamdef}) is in fact cryptoreal. Consider for simplicity only the case
(\ref{Vtild}). We have \footnote{Note that this Hamiltonian is not ${\cal PT}$-, but just ${\cal T}$-symmetric. },
 \be
H \ = \ \frac {p^2}2 + i\lambda p x^2 - i\beta p^3 - 2\lambda  x \bar \psi \psi \ ,
 \ee
where $ \beta = \lambda c^2/12$.

It is convenient to treat $\lambda$ and $\beta$ on equal footing and to get rid of
the complexities $\sim ipx^2$ and
$\sim i p^3$ simultaneously. The operator $R$ doing this job is
 \be
\label{RSQM}
R  \ =\ - \frac {\lambda x^3}3 + \beta x p^2 - 2 \lambda \beta x^2 \bar \psi \psi + \ldots \ ,
 \ee
where the dots stand for the terms of the third and higher order in $\lambda$ and/or $\beta$. The conjugated
Hamiltonian is
 \be
\label{hSQM}
\tilde{H} = e^R H e^{-R} \ =\ \frac {p^2}2 - 2\lambda x \bar\psi \psi + \frac 12 [ \lambda^2 x^4 + 3\beta^2 p^4]
+ \frac{1}{2}\lambda\beta + O(\lambda^3,\beta^3, \lambda^2 \beta, \lambda \beta^2)\,.
 \ee
It is Hermitian.
The rotated supercharges are
 \be
\label{QHerm}
 \tilde Q = e^R Qe^{-R} = \psi[p - i(\lambda x^2 - \beta p^2) + \lambda \beta x^2 p - \beta^2 p^3
 + \ldots]\,, \nonumber \\
\tilde {\bar Q} = e^R \bar{\cal Q} e^{-R} = \bar\psi [p + i(\lambda x^2 - \beta p^2)  + \lambda \beta x^2 p +
3\beta^2 p^3 + \ldots ]\,.
  \ee
We observe that they are still not adjoint to each other. To make them mutually adjoint
to the considered order in
$\beta, \lambda\,$, one should add to the operator $R$ one more term
\be
R \; \Rightarrow \; \hat{R} = R - 2\beta^2p^2\bar\psi \psi\,.
\ee
It is easy to see that this modification does not change the rotated Hamiltonian in the considered order,
but ensures the rotated supercharges to be manifestly adjoint to each other
 \be
\label{QHerm1}
 \hat{Q} = e^{\hat{R}} Qe^{-\hat{R}} = \psi[p - i(\lambda x^2 - \beta p^2) + \lambda \beta x^2 p
 + \beta^2 p^3 + \ldots]\,, \nonumber \\
\hat{\bar Q} = e^R \bar{\cal Q} e^{-R} = \bar\psi [p + i(\lambda x^2 - \beta p^2)  + \lambda \beta x^2 p +
\beta^2 p^3 + \ldots ]\,.
  \ee
By construction, the operators $\hat{Q}, \hat{\bar Q}$ and $\tilde{H}$ satisfy the standard
algebra (\ref{algebra}). We see that the requirement of the mutual adjointness of supercharges
is to some extent more fundamental than that of the Hermiticity of the Hamiltonian ---
the latter does not strictly fix the rotation operator $R$ while the former does.

 One can be convinced, order by order in $\beta, \lambda\,$, that complexities in $H$ can
be successfully rotated away also in higher orders (with simultaneously restoring the mutual conjugacy
of the supercharges), and this is also true for higher powers $N > 3$ in $V(X) \sim X^N$ and
hence for any analytic superpotential \footnote{It would be worth being aware of the full analytic
proof of this.}.

As the last topic of this Section, we shall consider NAC deformations of some other ${\cal N}=1$
SQM models \footnote{We denote by ${\cal N}$ the number of  complex
supercharges.}.

Besides the ${\cal N}=1$ multiplet with the off-shell content ${\bf (1, 2, 1)}$, there also exist
chiral ${\cal N}=1$ multiplets ${\bf (2,2, 0)}$ and ${\bf (0,2,2)}$,
having, correspondingly, even and odd overall Grassmann parity. They are described, respectively,
by the chiral superfields $\Phi(\theta, \bar\theta, t)$ and $\Psi(\theta, \bar\theta, t)$:
\be
\bar D \Phi = 0 \quad \Rightarrow \quad \Phi = z(t) + \theta \chi(t)\,, \quad
\bar D \Psi = 0 \quad \Rightarrow \quad \Psi = \omega(t) + \theta h(t)\,,
\ee
where as before $t = \tau - i\theta \bar \theta$, $z$ is a complex bosonic field, $\xi$ and $\omega$
are complex fermionic fields and
$h$ is a complex bosonic auxiliary field. It was shown in \cite{shap} that the only NAC deformation
preserving the $1D$ chirality and anti-chirality corresponds to the choice $\bar{C}=\tilde{C} = c^2 = 0,
C\neq 0$ in \p{deform}. Then the action of $\Phi$,
\be
S_{\Phi} = -\int dt \, d^2\theta \left[\frac{1}{4} D\Phi \bar D\bar \Phi + K(\Phi, \bar\Phi)\right]
= \int dt \left( \dot z \dot{\bar z} + \frac{i}{2}\bar\chi\dot\chi + \ldots \right), \label{Phi}
\ee
remains undeformed after replacing all products by the relevant $\star $ products \cite{shap}.

Actually, the same is true for the action of $\Psi $
\be
S_{\Psi} = -\int dt \, d^2\theta \left[ \frac{1}{4}\Psi\bar\Psi + \beta \bar\theta \Psi
- \bar\beta \theta \bar\Psi \right]  = \int dt \left(\frac{i}{2}\bar\omega\dot\omega
-\frac{1}{4}h\bar h + \beta h +
\bar\beta \bar h\right), \label{Psi}
\ee
where $\beta $ is a complex constant. However, while considering mutual couplings of
$\Psi$ and $\Phi$, there arise new possibilities. Prior to switching on any deformation, such
couplings provide potential terms for the ${\bf (2,2,0)}$ multiplet which do not exist
within the pure $\Phi$ system (the ``potential'' term $K(\Phi, \Bar\Phi)$ in \p{Phi}
produces only a Wess-Zumino type term $\sim (\dot z \bar z - \dot{\bar z} z) + \ldots \,$.).
In particular, one can consider the action
 \be
 \label{mixed}
S_{\Phi+\Psi} = -\int dt \, d^2\theta \left[\frac{1}{4} D\Phi \bar D\bar \Phi +
 \frac{1}{4}\Psi\bar\Psi + \bar \theta \Psi {\cal F}(\Phi) -
\theta \bar \Psi \bar {\cal F}(\bar \Phi) \right],
\ee
which  gives rise to a non-trivial
scalar potential for $z, \bar z$ upon eliminating the auxiliary fields $h, \bar h$ by
their equations of motion. For instance, choosing ${\cal F} = a + b\Phi + d \Phi^2$, one obtains  after
elimination of $h, \bar h$ the following on-shell component action
  \be
 S_{\Phi + \Psi} = \int dt \left[ |\dot z|^2 + \frac{i}{2}(\bar\chi\dot\chi + \bar\omega\dot\omega)
 + 4|a + b z + d z^2|^2 + \mbox{Yukawa $\omega, \chi$ couplings} \right].
  \ee
Nevertheless, once again, the direct (anti)chirality-preserving deformation of \p{mixed}
does not yield nothing new. The reason is that the terms proportional to the deformation
constant $C$  either never appear (in the antiholomorphic part $\sim \bar\Psi$) or vanish after doing
the Berezin integral (in the holomorphic part $\sim \Psi\,$).

There still exists an interesting mechanism of generating new potential terms via
the deformation. It is based on the observation that, while $\Psi^2= 0$ because of the Grassmann
character of $\Psi$, this nilpotency property is not longer valid for $\Psi\star\Psi$ and
higher-order star products. Indeed, we find
\be
\Psi\star \Psi = \frac{C}{2} h^2\,, \; \Psi\star(\Psi\star \Psi) = \frac{C}{2}\left(\omega h^2
+ \theta h^3\right), \; \Psi\star(\Psi\star \Psi\star \Psi) = \frac{C^2}{4} h^4\,, \quad \mbox{etc\,}.
\ee
The star products of $\bar \Psi$ coincide with the ordinary ones and so are identically zero.
Let us then e.g. add to the Lagrangian in \p{mixed}, with ${\cal F} = a + b z$ as the simplest choice,
the term $ {a_1}\,\bar\theta \Psi\star(\Psi\star \Psi)\,$. The bosonic
part of the corresponding component Lagrangian  is given by the following expression
\be
L = |\dot z|^2 -\frac{1}{4}h\bar h + h(a + bz) + \bar h (\bar a + \bar b\bar z) - \frac{a_1 C}{2} h^3\,.
\ee
Here we cannot longer treat $\bar h$ as a conjugate of $h$: both these fields should now be treated as
independent complex ones. Eliminating $\bar h$ by its equation of motion, we obtain
\be
L = |\dot z|^2 + 4|a + b z|^2 - 32 a_1 C (\bar a + \bar b \bar z)^3\,.
\ee
The additional term is holomorphic; by the same token as in Section 2 we conclude that the corresponding
term in the quantum Hamiltonian can be rotated away without trace! So the modified system proves to be
physically equivalent to the undeformed system (has the same quantum spectrum)
in spite of an apparent difference in their Lagrangians.

The star product deformation breaks a half of supersymmetries and the modified action is
manifestly invariant only under the holomorphic half of the original supersymmetry. Since after rotation
we reproduce the original system, the modified system should also respect some additional hidden
supersymmetry of the opposite holomorphy, like in the ${\bf (1, 2, 1)}$ system of Ref.
\cite{shap} discussed above.

\section{Field theories}

The first example of an anticommutative deformation of a supersymmetric field theory was considered in
Ref.\cite{Seiberg}. Seiberg took the standard Wess-Zumino model
 \be
\label{WZ}
{\cal L} \ =\ \int d^4\theta \, \bar \Phi \Phi + \left[ \int d^2\theta \left( \frac {m\Phi^2}2
+ \frac {\lambda \Phi^3}3
\right) + {\rm c.c} \right]
 \ee
(where now $\int d^2\theta\, (\theta^\alpha \theta_\alpha) = 1\,$) and deformed it by introducing the nontrivial anticommutator
  \be
\label{Calbet}
 \{\theta^\alpha, \theta^\beta \} \ =\ C^{\alpha\beta}\ ,
 \ee
$  C^{\alpha\beta} =  C^{\beta\alpha}$, in the assumption that all other (anti)commutators vanish,
  \be
\label{drugiekom}
\{\bar \theta^{\dot \alpha}, \bar \theta^{\dot \beta} \} = \{ \theta^{\alpha}, \bar \theta^{\dot \beta} \} =
[\theta^\alpha, x^L_\mu] = [ \bar \theta^{\dot \alpha}, x^L_\mu] = [x^L_\mu, x^L_\nu] = 0\ .
  \ee
Note that this all was written in the {\it chiral} basis,
$x_\mu^L = x_\mu^{\rm central} + i\theta \sigma_\mu \bar \theta $.
 In Ref.\cite{Seiberg}, the space $x_\mu$ was assumed to be Euclidean. We will work in Minkowski space, however,
 and will not be scared
by the appearance of complexities  at intermediate steps.  The Minkowski space deformation
(\ref{Calbet}), (\ref{drugiekom}) is analogous
to the SQM deformation (\ref{deform}) with $\bar C = \tilde C = 0\,$.

The anticommutator
(\ref{Calbet}) introduces a constant self-dual tensor which explicitly breaks Lorentz invariance.
However, the deformed Lagrangian expressed in terms of the component fields proves still
to be Lorentz invariant. Indeed, it is easy to find that the
 kinetic term $\int d^4\theta \, \bar \Phi \Phi $ is undeformed and the only extra piece comes from
 \be
 \label{F3}
\frac \lambda 3 \int d^2\theta \, \Phi^3 \ \to\   \int d^2\theta \, \Phi*\Phi*\Phi =
 F(m\phi + \lambda \phi^2)
-   \frac \lambda 3  \,\det \|C\| F^3\ .
 \ee
 It depends only on
the scalar $~\det\|C\|$ and is obviously Lorentz invariant. Adding
$ \bar F(m \bar\phi + \lambda \bar \phi^2)$
from $\int d^2 \bar\theta \bar\Phi^3$ and $F\bar F$ from the kinetic term, and
expressing $F$ and $\bar F$ via
$\phi$ and $\bar\phi$, we see that the undeformed potential $|m\phi + \lambda \phi^2|^2$
acquires an extra holomorphic contribution $\sim (m\bar\phi + \lambda \bar\phi^2)^3$.

We have seen, however, that such a holomorphic deformation can be rotated away without
trace! In other words, the deformation (\ref{Calbet}) does not change the dynamics
(the spectrum of the Hamiltonian etc) of the Wess-Zumino model in Minkowski
space \footnote{To avoid a misunderstanding, we
would like to point out that even in Minkowski space, the fields $\phi$ and $\bar\phi$ (as well
as $F$ and $\bar F$) after deformation should be treated as complex fields which {\it are not conjugate}
to each other. The standard complex conjugacy requirements can be consistently imposed
on the {\it rotated} fields and their canonical momenta.}.

 The final example is the deformed ${\cal N} = 2$ gauge theory \cite{N2,2N2}. There exists
 in this case a natural Lorentz invariant
deformation \cite{N2},
  \be
\label{N2deform}
\{\theta^\alpha_i, \theta^\beta_j \}\ =\ \frac{1}{4} J \epsilon^{\alpha\beta} \epsilon_{ij}\ ,
 \ee
$i,j = 1,2$ \footnote{The deformation parameter $J$ is related to the original one $I$ \cite{N2} as
$J = 4I\,$.}. The Lagrangian of the deformed ${\cal N} = 2$ supersymmetric $U(1)$ theory is \cite{N2,2N2}
  \be
&& {\cal L} = {\cal L}_{\phi} + {\cal L}_{\Psi} + {\cal L}_{A}\,, \label{N2defAct} \\
&& {\cal L}_{\phi} = -\frac{1}{2}\Box \bar\phi\left[\phi + \frac{J A_mA_m}{1 + J\bar\phi} +
\frac{1}{4}\frac{J^3 \partial_m\bar\phi\partial_m\bar\phi}{1 + J\bar\phi} \right], \label{phiAct}\\
&&{\cal L}_\Psi = i\left[\Psi^{i\alpha} + \frac {J A_m (\sigma_m)^\alpha_{\dot\alpha}
\bar \Psi^{i\dot \alpha}}{1 + J\bar \phi} \right] (\sigma_n)_{\alpha\dot\beta} \, \partial_n
\left( \frac {\bar \Psi_i^{\dot\beta}}{1 + J\bar\phi} \right),
\label{PsiAct} \\
&&{\cal L}_{A} = \frac{1}{4}(1 + J\bar\phi)^2\left(f_{mn}f_{mn}
+ f_{mn}\tilde{f}_{m n}\right), \label{AAct} \\
&& f_{mn} = \partial_m \left(\frac{1}{1 + J\bar\phi} A_n\right)-
\partial_n \left(\frac{1}{1 + J\bar\phi} A_m\right).
\ee

The Lagrangian \p{N2defAct} was derived originally in Euclidean space.
 In Minkowski space, it is clearly complex. Bearing in mind the previous discussion, it is natural
to suggest, however, that the corresponding Hamiltonian is cryptoreal.
Leaving the issue of cryptoreality of the full field theory Hamiltonian for  future study, let us disregard
the fermion part of \p{N2defAct} and consider the
$1D$ reduction of what is left. We will  show that the resulting quantum-mechanical model is cryptoreal
and actually amounts to the free model. The reduction goes as
\be
\Box \to \partial^2_t\,, \; \partial_m\phi\partial_m\bar\phi \to \dot\phi\dot{\bar\phi}\,, \;
A_nA_n \to -A_0A_0 + \vec{A}\vec{A}
\ee
and we obtain
\be
{\cal L}_{bos}^{QM} = \frac{1}{2}\dot{\phi}\dot{\bar\phi} + \frac{1}{2}\dot{\vec{A}}\dot{\vec{A}}
-\frac{1}{24}\frac{J^4(\dot{\bar\phi})^4}{(1 + J\bar\phi)^2}
+ \frac{J}{2} \frac{\ddot{\bar\phi}}{1 + J\bar\phi} A^2_0\,.
\ee
The corresponding canonical Hamiltonian, in the obvious notation, is as follows
\be
H = 2P \bar P + \frac{1}{2}\, \vec{P}\vec{P}+ \frac{2}{3}J^4\frac{P^4}{(1 + J\bar\phi)^2} - 2J^2 A^2_0
\frac{P^2}{(1 + J\bar\phi)^2}\,.
\ee
Making the rotation
\be
H' = e^R H e^{-R}\,, \quad R = -\frac{i}{3}\frac{J^3 P^3}{1 + J\bar\phi} + iA^2_0 \frac{J P}{1 + J\bar\phi}\,,
\ee
we find that
\be
H' = 2P \bar P + \frac{1}{2}\, \vec{P}\vec{P}\,,
\ee
i.e. the deformation is rotated away without trace, like in the examples above, and
our quantum-mechanical system is reduced to the free one. In the full 4-dimensional case
the situation is more subtle due to the presence of the term $\sim \varepsilon_{mnrq}$ that vanished after
reduction. Our simple $1D$ consideration shows that the corresponding dynamics
in Minkowski space is expected to be ``almost trivial''. Nevertheless, we do not see reasons
why the deformation in this case can be entirely rotated away. Rather, the situation should be similar
to what we observed in the Aldrovandi-Schaposnik model. To get a deeper insight into these issues,
it would be instructive to analyze, from a similar point of view, the deformations of
the nonabelian ${\cal N}=2$ gauge theories \cite{N2} and the models involving
hypermultiplets \cite{N22}, which are not free in the undeformed limit $J=0\,$.

\section{Discussion}
 Our main result is that NAC deformations of supersymmetric theories are well defined
not only in Euclidean, but also in Minkowski space. In spite of its unfriendly looking complex
appearance, the deformed theory can be endowed with a Hilbert space where the Hamiltonian is
Hermitian and its spectrum is real. In many cases (in particular in the case of the deformed Wess-Zumino
model considered in Seiberg's original paper), the deformed Hamiltonian is actually physically equivalent
to the undeformed one. Extra contributions stemming from nonanticommutativity have holomorphic structure
and can be ``rotated away'' without trace, as was explained in the text of the paper. For some other NAC
theories, the new Hamiltonian is not equivalent to the old one and deformation brings about nontrivial
changes in dynamics.

We discussed at length a one-dimensional SQM example due to Aldrovandi and Schaposnik. While going to
$4D$ field theories, the requirement that Lorentz invariance is kept after deformation dictates
the undeformed theory
to possess at least  ${\cal N} = 2$ supersymmetry [see Eq.(\ref{N2deform})]. The Lagrangian
of the deformed ${\cal N} = 2$ gauge theory
was constructed before. We have not proven, but argued that it is cryptoreal (i.e. the Hamiltonian
can be made Hermitian) but is not equivalent to the undeformed Lagrangian.
A thorough study of this interesting question is a problem for the future.

Another interesting direction of study, not related to nonanticommutativity, but related to cryptoreality
is the following. In Ref. \cite{ISZ}, we constructed a gauge theory in six dimensions which is
superconformal at the classical level. It is renormalizable, and the variant of the theory
involving interaction with a hypermultiplet \cite{hyp} is anomaly free \cite{anom}. However,
this theory involves higher derivatives, which may in principle lead to the  loss of unitarity due to the
presence of ghosts. In particular, the theory involves scalar fields $D$ of canonical dimension 2
with the potential $\sim D^3$. Naively, such a potential means vacuum instability and the associated
loss of unitarity. We have seen, however, that the QM models with the potentials $V(x) \sim ix^3$
or $V(x) \sim x^3$ {\it can} be meaningful since their Hamiltonians can be made Hermitian.
It is not excluded that this is also the case for certain higher-derivative field theories and,
in particular, for the models constructed in \cite{ISZ,hyp}.

\section*{Acknowledgements}

E.I. acknowledges a support from RFBR grant, project  No
06-02-16684, NATO grant PST.GLG.980302, the grant INTAS-05-7928,
the FRBR-DFG grant 06-0204012, the DFG grant No 436 RUS 113/669/0-3 and a grant of the
Heisenberg-Landau program. He thanks B. Zupnik for useful discussions and the SUBATECH for
the warm hospitality in Nantes. A.S. is grateful to C. Bender for the interest in the work
and valuable correspondence.

\end{document}